%% file: das.tex
\documentclass[proceedings]{stacs}
\stacsheading{2010}{215-226}{Nancy, France}
 \firstpageno{215}

\usepackage{algorithmic}
\usepackage{algorithm}
\usepackage{xcolor}
\usepackage{amssymb}
\usepackage{latexsym}
\usepackage{amsmath}

\newenvironment{appendix-theorem}[1]{\vspace{\theorempreskipamount}\noindent{\bf Theorem~#1~} \em }{\vspace{\theorempostskipamount}}
\newenvironment{appendix-lemma}[1]{\vspace{\theorempreskipamount}\noindent{\bf Lemma~#1~} \em }{\vspace{\theorempostskipamount}}
\newenvironment{appendix-corollary}[1]{\vspace{\theorempreskipamount}\noindent{\bf Corollary~#1~} \em }{\vspace{\theorempostskipamount}}
\newenvironment{appendix-claim}[1]{\vspace{\theorempreskipamount}\noindent{\bf Claim~#1~} \em }{\vspace{\theorempostskipamount}}

\newcommand{\NL}{\mbox{{\sf NL}}}
\newcommand{\SPL}{\mbox{{\sf SPL}}}
\newcommand{\FL}{\mbox{{\sf FL}}}
\newcommand{\SL}{\mbox{{\sf SL}}}

\newcommand{\Ptime}{\mbox{{\sf P}}}
\newcommand{\RNC}{\mbox{{\sf RNC}}}
\newcommand{\NC}{\mbox{{\sf NC}}}
\newcommand{\UL}{\mbox{{\sf UL}}}

\newcommand{\Log}{\mbox{{\sf L}}}

\newcommand{\LogCFL}{\mbox{{\sf LogCFL}}}

\newcommand{\NLC}{\mbox{{\sf NL-complete}}}

\newcommand{\Reach}{\mbox{{\sf Reach}}}
\newcommand{\Distance}{\mbox{{\sf Distance}}}
\newcommand{\LongPath}{\mbox{{\sf Long-Path}}}

\newcommand{\Class}{\mbox{${\mathcal C}$}}
\theoremstyle{plain}
\theoremstyle{definition}

\newtheorem{claim}[thm]{Claim} 
\begin{document}

% \title[short title]{title}
%\title[Using stacs.cls]{How to use stacs.cls}
\title[Log-space Algorithms for Paths and Matchings in $k$-trees]{Log-space Algorithms for Paths and Matchings in $k$-trees}

% \author[ref]{Short author}{Author}
\author[imsc]{B. Das}{Bireswar Das}
% \address[ref]{Address of authors with ref as reference}
\address[imsc]{The Institute of Mathematical Sciences
  \newline Chennai, India}  %required
\email{{bireswar, prajakta}@imsc.res.in}  %optional

\author[cmi]{S. Datta}{Samir Datta}
\address[cmi]{Chennai Mathematical Institute, Chennai, India}	%optional
\email{sdatta@cmi.ac.in}  %optional

\author[imsc]{P. Nimbhorkar}{Prajakta Nimbhorkar}

%% mandatory lists of keywords and classifications:
\subjclass{Computational Complexity}
% \titlecomment{OPTIONAL comment concerning the title, \eg, if a variant
% or an extended abstract of the paper has appeared elsewehere}
%%%%%%%%%%%%%%%%%%%%%%%%%%%%%%%%%%%%%%%%%%%%%%%%%%%%%%%%%%%%%%%%%%%%%%%%%%%

%% the abstract has to PRECEDE the command \maketitle:
%% be sure not to issue the \maketitle command twice!

\begin{abstract}
Reachability and shortest path problems are \NLC\ for general graphs. They
are known to be in \Log\ for graphs of tree-width $2$ \cite{JT07}. However,
for graphs of tree-width larger than $2$, no bound better than \NL\ is known.
In this paper, we improve these bounds for $k$-trees, where $k$ is a constant. In 
particular, the main results of our paper are log-space algorithms for reachability in directed $k$-trees, 
and for computation of shortest and longest paths in directed acyclic $k$-trees.
%shortest path computation in undirected $k$-trees, $k+1$-coloring of $k$-trees
%and for some perfect matching problems in $k$-trees. 

Besides the path problems mentioned above, we consider the problem of deciding
whether a $k$-tree has a perfect macthing (decision version), and if so, finding
a perfect matching (search version), and prove that these problems are \Log-complete. 
These problems are known
to be in \Ptime\ and in \RNC\ for general graphs, and in \SPL\ for planar bipartite
graphs \cite{DKR08}. 

Our results settle the complexity of these problems for the class of $k$-trees. The results are also applicable
for bounded tree-width graphs, when a tree-decomposition is given as input. The technique
central to our algorithms is a careful implementation of divide-and-conquer approach in log-space, along with
some ideas from \cite{JT07} and \cite{LMR07}.
\end{abstract}
\maketitle

\input{1-intro.tex}
\input{2-prelim.tex}
\input{3-reach.tex}
\input{4-shortlong.tex}
\input{5-matchings.tex}
\bibliography{das}

\bibliographystyle{plain}
\end{document}

%% file: 1-intro.tex
\section{Introduction}
Reingold's striking result \cite{Reingold}, showed that undirected reachability
is in \Log, thus collapsing the class \SL\ to \Log. On the other hand,
directed reachability, which happens to be \NLC\  is another similar sounding problem
for which there is only partial progress to report. A result of Allender and Reinhardt,
\cite{RA97} hints at a partial collapse of \NL\ by showing that directed reachability
is in the formally smaller class \UL, although, \emph{non-uniformly}.

In the absence of better constructive upper bounds it is natural to consider 
natural restrictions on graphs which allow us to improve the upper bounds
on reachability and related problems. Typical examples of this approach
are \cite{ABCDR09},\cite{TW09}, where the complexity of various versions of
planar and somewhat non-planar (in the sense of excluding only a $K_5$
or only a $K_{3,3}$ minor) are considered. In the same spirit, but using 
different techniques, \cite{JT07} considers reachability and related
questions in series-parallel graphs and places all of these in \Log.
They leave open the question of complexity of such problems in 
bounded tree-width graphs. 
Series-parallel graphs have tree-width two and happen to be
planar. But higher tree widths graphs are highly non-planar.
In fact, any $k$-tree for $k > 4$ contains both $K_5$ and $K_{3,3}$.

We resolve the open questions posed in \cite{JT07} and show a matching
\Log\ lower bound to complete the characterization of reachability problems
in $k$-trees. Thus one of the main results of our paper is the following:
\begin{theorem}\label{thm:reach}
The following problems are \Log-complete:

%\begin{enumerate}
1. Computing reachability between two vertices in directed $k$-trees,

2. Computing shortest and longest paths in directed acyclic $k$-trees.
%\end{enumerate}
\end{theorem}
In this paper, we also consider the perfect matching problem.
The parallel complexity of perfect matching problems is a long standing
open problem where the best known algorithms use randomness as a resource
\cite{MVV87},\cite{KUW86}. Even in the planar case, the search problem for
perfect matchings is known to be in \NC\ for bipartite graphs only
\cite{DKR08}.

We prove a complete characterization for the decision and search
versions of the perfect matching problem for $k$-trees.
This improves significantly upon previous best known
upper bound of \LogCFL\ for bounded tree-width graphs. 
Thus another main result of our paper is:
\begin{theorem}\label{thm:match}
Deciding whether a $k$-tree has a perfect matching, and if so, finding a perfect matching 
is \Log-complete.
\end{theorem}

Our primary technique is a careful use of divide-and-conquer to enable
the algorithm to run in \Log. However, for the distance computation
we need to import a constructive version of tree separation from
\cite{LMR07} where it is stated in the context of Visibly Pushdown Automata (VPAs).
We believe that porting this technique for use in general log-space computation
is an important contribution of this paper.

At this point, we must mention an important caveat. All our log-space
results hold directly only for $k$-trees and not for partial $k$-trees
which are also equivalent to tree-width $k$ graphs. The reason being
that a tree decomposition for partial $k$-trees is apparently more
difficult to construct (best known upper bound is \LogCFL \cite{Wanke94}) 
as opposed to $k$-trees (for which it can be done in \Log\ \cite{KK09}). Having mentioned that
it is important to observe that if we are given the tree decomposition
of a partial $k$-tree, we can do the rest of computation in \Log.

The rest of the paper is organized as follows: Section \ref{sec:prel} gives the necessary background. Section \ref{sec:reach} contains
log-space algorithms for reachability in directed $k$-paths and $k$-trees. 
Section \ref{sec:path} contains log-space algorithms for shortest and longest path 
in directed acyclic $k$-paths and $k$-trees. Section
\ref{sec:mat} contains log-space algorithms for perfect matching problems in a $k$-tree. 

%% file: 2-prelim.tex
\section{Preliminaries}\label{sec:prel}
We define $k$-trees and a subclass of $k$-trees known as $k$-paths here, 
and also describe a suitable representation for the graphs in these two classes. This representation is
used in our algorithms in the rest of the paper.
All the definitions given here are applicable to both directed as well as undirected graphs.
For directed graphs, the directions of the edges can be ignored while defining $k$-trees and $k$-paths and while
computing their suitable representations.

The class of graphs known as $k$-trees is defined as (cf. \cite{HP68} ):
\begin{definition}
  The class of \emph{$k$-trees} is inductively defined as follows.
\begin{itemize}
\item A clique with $k$ vertices ($k$-clique for short) is a $k$-tree.
\item Given a $k$-tree $G'$ with $n$ vertices, a $k$-tree $G$ with
  $n+1$ vertices can be constructed by 
  picking a $k$-clique $X$ (called the \emph{support})in $G'$ and then
  joining a new vertex $v$ to each
  vertex $u$ in $X$. Thus, $V(G)=V(G')\cup\{v\}$,
  $E(G)=E(G')\cup\{\{u,v\}\mid u\in X\}$.
\end{itemize}

\end{definition}

A \emph{partial $k$-tree} is a subgraph of a $k$-tree. 
The class of partial $k$-trees
coincides with the class of graphs that have tree-width at most $k$.
$k$-trees are recognizable in log-space \cite{ADK07}
but partial $k$-trees are not known to be recognizable in log-space. 
% A special subclass of $k$-trees is $k$-path. 
% \begin{definition}
%   An \emph{interval graph} is a graph whose vertices can be put in one
%   to one correspondence with a set of open intervals on the real line
%   such that two vertices are adjacent if and only if the corresponding
%   intervals have a nonempty intersection.
% \end{definition}
% \begin{definition}{\rm\cite{Klo94}}
%   A \emph{$k$-path} is a $k$-tree which is an interval graph.
% \end{definition}
In literature, several different
representations of $k$-trees have been considered 
\cite{GSS02,ADK07,KK09}. We use the following representation given by K\"obler and Kuhnert
\cite{KK09}: 
\begin{definition}
Let $G=(V,E)$ be a $k$-tree. The tree representation $T(G)$ of $G$ is
defined by
$$V(T(G))=\{M\subseteq V\mid M \textrm{ is a $k$-clique or a
$(k+1)$-clique}\}, \\$$
$$E(T(G))=\{\{M_1,M_2\}\subseteq V\mid M_1\subsetneq M_2\}$$
\end{definition}
%The following theorem appears in \cite{KK09}:
%\begin{theorem}\label{trrep}
%Let $G$ be a $k$-tree. Then
%\begin{itemize}
%\item The tree representation $T(G)$ of $G$ is a tree.
%\item The tree representation $T(G)$ can be computed in $\FL$.
%\end{itemize}
%\end{theorem}
In \cite{KK09}, it is proved that $T(G)$ is a tree and can be computed in
log-space. In the rest of the paper, we use $G$ in place of $T(G)$. Thus, by \textit{a $k$-tree $G$}, we 
always mean that $G$ is in fact represented as $T(G)$. The term \textit{vertices in $G$} refers to the vertices
in the original graph, whereas \textit{a node in $G$} and \textit{a clique in $G$} refer to the nodes of $T(G)$.
Partial $k$-trees also have a tree-decomposition similar to that of $k$-trees,
which is also not known to be log-space computable. 

\textit{$k$-paths} is a sub-class of $k$-trees (e.g. see \cite{GNPR05}). 
The recursive definition of $k$-paths is similar to
that of $k$-trees. However, a new vertex can be added only to 
a particular clique called the \emph{current clique}. After addition of a vertex,
the current clique may remain the same, or may change by
dropping a vertex and adding the new vertex in the current clique.
We consider the following representation of $k$-paths, which is based on
the recursive definition of $k$-paths, and is known to be computable in log-space \cite{ADK07}:

%\begin{definition}
Given a $k$-path $G=(V,E)$, for $i=1,\cdots,m$, let $X_i$ be the current cliques  at the $i$th stage of  the recursive construction of the $k$-path.  
% Thus, $X_i$ is the $i$th current $k$-clique.
Let
$V_1=\cup_{i}X_i$ and $V_2=V\setminus V_1$. We call the vertices in $V_2$ as \textit{spikes}.
%Each clique possibly has
%singleton vertices connected to the entire clique by edges. We call such
%vertices as \textit{spikes}.
%Vertices in $V_2$ do not appear in any of the $X_i$'s but they are connected to
%all the vertices in exactly one of the $X_i$'s. 
%Also, no two vertices in $V_2$ have
%an edge between them. We call such vertices as \textit{spikes}. 
%In $G$, $V_2$ is the set of all such spikes, and each spike
%is connected to exactly one of .  
The following facts are easy to see:
%\begin{enumerate}

1. No two spikes have an edge between
them. 
%We call this decomposition as $k$-path decomposition, and it can be
%computed in log-space (cf. \cite{ADK07}).  

2. Each spike is connected to all the vertices of exactly one of the $X_i$'s.

3. $X_i$ and $X_{i+1}$ share exactly $k-1$ vertices
%\end{enumerate}

The representation of $G$ consists of a graph $G'=(V',E')$ where
$V'=\{X_1,\ldots,X_m\}\cup V_2$ and $E'=\{(X_i,X_{i+1}) \mid  1\leq i<m\}\cup \{(X,v)|X \textrm{ is a clique in }\in V', v\in V_2 \textrm{ has a neighbour in } X\}$.

%% file: 3-reach.tex
\section{Reachability} \label{sec:reach}
We give log-space algorithms to compute reachability in $k$-paths and
in $k$-trees. Although the graphs considered in this section are
directed, when we refer to any of the definitions or decompositions in 
Section \ref{sec:prel}, we consider the underlying undirected graph. 
%A $k$-path is a $k$-tree 
%which is an interval graph. We recall that 
%an \emph{interval graph} is a graph whose vertices can be put in one
%  to one correspondence with a set of open intervals on the real line
%  such that two vertices are adjacent if and only if the corresponding
%  intervals have a nonempty intersection.
%\end{definition}
%\begin{definition}{\rm\cite{Klo94}}
%  A \emph{$k$-path} is a $k$-tree which is an interval graph.
%\end{definition}
\subsection{Reachability in $k$-paths}\label{sec:kpreach}
% Given two designated vertices $s$ and $t$ in $G$, we check whether it
% has a $k$-path decomposition, and if so, check whether $t$ is reachable
% from $s$ in log-space. Here onwards, by $G$, we refer to a k-path
% decomposition of $G$.
Without loss of generality, we can assume that $s$ and $t$ are vertices
in some $k$-cliques $X_i$ and $X_j$, and not spikes. If $s$ ($t$) is a
spike, then it has at most $k$ out-neighbors (resp. in-neighbors)
and we can take one of the out-neighbors (resp. in-neighbors) as the
new source $s'$ and new sink $t'$ and check reachability.
As there are only $k^2$ such pairs, we can cycle through all of them in
log-space. The algorithm is based on the observation that a simple $s$ to $t$ path
$\rho$ can pass through any clique at most $k$ times. We use a divide-
and-conquer approach similar to that used in Savitch's algorithm (which
shows that directed reachability can be computed in $DSPACE(\log^2{n})$). 
The main steps involved in the algorithm are as follows:
%\begin{enumerate}

1. Preprocessing step: Make the cliques disjoint by labeling
different copies of each vertex with different labels and introducing
appropriate edges. Compute reachabilities within each clique including
its spikes, and \emph{remove the spikes}. Number the cliques
$X_1,\ldots, X_m$ left to right.

2. Now assume that $s$ and $t$ are in cliques $X_i$ and $X_j$
respectively. Note that $i=j$ is also possible, but without loss of
generality, we can assume $i<j$. This is because, if $i=j$, we can make
another copy $X_i'$ of $X_i$, join the copies of the same vertex by
bidirectional edges to preserve reachabilities, and choose the copy 
of $s$ from $X_i$ and that of $t$ from $X_i'$.

3. Divide the $k$-path into three parts $P_1,~P_2$ and $P_3$ where $P_1$ 
consists of cliques $X_1,\ldots,X_i$, $P_2$ consists of $X_i,\ldots,X_j$, 
and $P_3$ consists of $X_j,\ldots,X_m$. Note that $X_i$ ($X_j$) appears in 
both $P_1$
and $P_2$ ($P_2$ and $P_3$ respectively). Now compute reachabilities
of all pairs of vertices in $X_i$ ($X_j$) when the graph is restricted to $P_1$
(respectively $P_3$). Then the reachability of $t$ from $s$ within $P_2$
is computed, using the previously computed reachabilities within $P_1$
and $P_3$. 
%\end{enumerate}

Each of these steps can be done by a log-space transducer.
The details are given below.

{\bf Preprocessing: } Although adjacent k-cliques in a k-path decomposition
share $k-1$
vertices, we perform a preprocessing step, where we give distinct labels
to each copy of a vertex. As all the copies of a vertex form a
(connected) sub-path in the k-path decomposition, we join two copies of a
vertex appearing in two adjacent cliques by bidirectional edges. It can
be seen that this preserves reachabilities. Any copy of $s$ and $t$ 
can be taken as the new $s$ and $t$. 
Another preprocessing step involves removing the spikes
maintaining reachabilities between all pairs of vertices, and 
computing reachabilities within each k-clique. 
Both of these preprocessing steps can be done by a 
log-space transducer. The proof appears in the full version of the
paper.
%\begin{lemma}\label{lem:spikes}
%The spikes attached to each clique can be removed in log-space, maintaining
%reachability between every pair of vertices in the rest of the $k$-path.
%\end{lemma}

{\bf The Algorithm: }
We describe an algorithm to compute pairwise reachabilities in $X_i$ and 
$X_j$ in $P_1$ and $P_3$ respectively, and also $s$-$t$ reachability in
$P_2$ using these previously computed pairwise reachabilities. 
Algorithm \ref{alg:isreach} describes this reachability routine. The
routine gets as input two vertices $u$ and $v$, and two indices $i$
and $j$. It determines whether $v$ is reachable from $u$ in the sub-path
$P=(X_i,\ldots, X_j)$. This input is given in such a way that $u$ and $v$ 
always lie in $X_i$ or $X_j$. Consider the case when both $u$ and $v$
are in $X_i$ (or both in $X_j$). Let $l$ be the center of $P$. 
Then a path from $u$ to $v$ either lies entirely in the sub-path
$P'=(X_i,\ldots,X_l)$ or it crosses $X_l$ at most $k$ times. Thus if
$X_l=\{v_1,\ldots,v_k\}$ then for $\{v_{i_1},\cdots,v_{i_r}\}\subseteq
X_l$ we need to check reachabilities between $u$ and
say $v_{i_1}$ in $P'$, then between $v_{i_1}$ and $v_{i_2}$ in
$P''=(X_l,\ldots,X_j)$ and so on, and finally between $v_{i_r}$ and $v$
in $P'$.
It suffices to check all the $r$-tuples in $X_l$, where $0\leq r\leq k$.
The case when $u\in X_i$ and $v\in X_j$ (and vice versa) is analogous.
In Algorithm \ref{alg:isreach}, we present only one case where $u,v\in
X_i$. Other three cases are analogous.
Thus at each recursive call, the length of the sub-path under
consideration is halved, and $O(\log{m})$ iterations suffice. 
\begin{algorithm}[!ht]
\caption{Procedure IsReach($u$, $v$, $i$, $j$)}\label{alg:isreach}

\begin{algorithmic}[1]
	\STATE Input: Pre-processed k-path decomposition of graph $G$,
	clique indices $i,j$, vertex labels $u,v\in X_i$. \COMMENT{Other
	three cases are analogous.}
	\STATE Decide: Whether $v$ is reachable from $u$ in sub-path
	$P=(X_i,\ldots, X_j)$.
	\IF{$j-i=1$}
		\STATE Compute the reachability directly, as the sub-path has only $2k$
		vertices.
		\STATE Return the result.
	\ENDIF
	\STATE $l=\frac{j+i}{2}$
	\IF{$u,v\in X_i$}
		\IF{IsReach($u$, $v$, $i$, $l$)} 
			\STATE Return 1;
		\ELSE
			\FOR{$q=1$ to $k$}
				\STATE $v_0\leftarrow u$, $v_{q+1}\leftarrow v$
				\FORALL{$q$-tuples ($v_1,\ldots,v_q$) of
				vertices in $X_l$}
					\IF{$\bigwedge_{\substack{x=0
					\\x\textrm{ even}}}^{q+1}$
					IsReach($v_x$,$v_{x+1}$,$i$,$l$) 
					$\land$
					$\bigwedge_{\substack{x=1 \\ 
					x\textrm{ odd}}}^{q+1}$
					IsReach($v_x$,$v_{x+1}$,$l$,$j$)}
						\STATE Return 1;
					\ENDIF
				\ENDFOR
			\ENDFOR
		\ENDIF
	\ENDIF
\end{algorithmic}
\end{algorithm}
The algorithm can be implemented in log-space. The correctness and complexity analysis of the algorithm appears in the full version.

\subsection{Reachability in $k$-trees} \label{sec:ktreach}

Given a directed $k$-tree $G$ in its tree decomposition and two vertices $s$ and $t$ in $G$,
we describe a log-space algorithm that checks whether $t$ is reachable from $s$. 
%This implies 
%log-space algorithm for reachability
%in $k$-trees and graphs with tree-distance width bounded by $k$.
This algorithm uses Algorithm~\ref{alg:isreach} as a subroutine and
involves the following steps:
The complexity analysis is given in Lemma \ref{lem:ktrc}.
%Let $D=(T=(V_T,E_T),\{X_i\}_{V_T})$ be a tree-like decomposition of
%$G=(V,E)$ of width at most $k$. Let $s,t\in V$. The task is to decide
%if $t$ is reachable from $t$ in $G$.  
% Now we describe a log-space algorithm for computing reachability in
% k-trees, and graphs with tree-distance-width at most $k$, 
% where $k$ is a fixed constant. As described in Section \ref{sec:dec},
% the tree-decomposition of k-trees and k-tree-distance-width graphs can
% be computed in log-space. The reachability algorithm has following
% steps, each of which can be executed in log-space: 
%\begin{enumerate}

1. \textbf{Preprocessing: }
Like $k$-paths, assign distinct labels to the copies of each vertex $u$ in different cliques.
Introduce a bidirectional edge between the copies of
$u$ in all the adjacent pairs of cliques. As reachabilities are maintained during this process, any
copy of $s$ and $t$ can be taken as the new $s$ and $t$. Let $X_i$ and
$X_j$ be the cliques containing $s$ and $t$ respectively. 
%Note that each tree-edge is actually made up of $k$ or $k-1$
%bidirectional edges, since two adjacent cliques have $k-1$ vertices in
%common, and two adjacent copies of the same clique have $k$
%bidirectional edges between them. 

2. \textbf{The Procedure: }
After this preprocessing, we have a tree $T$ with its nodes
as disjoint $k$-cliques of vertices of $G$, and $s$ and $t$ are contained in cliques 
$X_i$ and $X_j$. Compute the unique undirected path 
$\rho$ between
$X_i$ and $X_j$ in $T$ in log-space. Each node on $\rho$ has two of its
 neighbors on $\rho$, except $X_i$ and $X_j$, which have one
neighbor each. An $s$ to $t$ path has to cross each clique in $\rho$,
and additionally, it can pass through the subtrees attached to each node
$X_l$ on $\rho$. 
Hence for each node $X_l$ on $\rho$, we pre-compute the pairwise reachabilities
among the $k$ vertices contained in $X_l$ when the $k$-tree is restricted to the subtree rooted at
$X_l$. We define the subtree rooted at $X_l$ as the subtree consisting of $X_l$ and 
those nodes which can be reached from $X_l$ without going through any node on $\rho$. 
%leaving its neighbors on $\rho$. 
Note that once this is done for each node $X_l$ on $\rho$, we are left with $\rho$. As $\rho$ is a $k$-path, we can
use Algorithm \ref{alg:isreach} in Section \ref{sec:kpreach} to compute reachabilities within $\rho$.
%So we can compute reachability between $s$ and $t$ in
%$\rho$ in log-space as mentioned in Remark~\ref{rem:path-like}.

3. \textbf{Computing reachabilities within the subtree rooted at $X_l$:}
We do this inductively. 
%As $T$ is a binary tree and $X_l$ has at least 
%one neighbor on $\rho$,
%it can have at most two children, say $X_a$ and $X_b$ in the subtree rooted at it. 
If the subtree rooted at $X_l$ contains only one
node $X_l$, we have only $k$ vertices, and their pairwise reachabilities
within $X_l$ can be computed in $O(k\log{k})$ space. 
%This holds even
%when $X_l$ has no non-leaf child. This is the base case of induction.
We recursively find the reachabilities within the subtrees rooted at each of the children of $X_l$.
%within the subtrees rooted at each of them. 
Let the size of the subtree rooted at $X_l$ be $N$. At most one of the children
of $X_l$ can have a subtree of size larger than $\frac{N}{2}$. 
Let $X_a$ be such a child. 
Recursively
compute the pairwise reachabilities for each pair of vertices in $X_a$ within the subtree rooted at $X_a$.
%, that is
%reachabilities among vertices of $X_b$ within the subtree rooted at
%$X_b$. 
The reachabilities are represented as a $k\times k$ boolean matrix referred to as the
\textit{reachability matrix} $M$ for the vertices in $X_a$, when the graph is confined to
the subtree rooted at $X_a$. $M$ is then used to compute the pairwise reachabilities of vertices
in $X_l$, when the graph is confined to $X_l$ and the subtree rooted at $X_a$. This gives a new
matrix $M'$ of size $k^2$. It is stored on stack while computing the reachability 
matrix $M''$ for another child $X_b$ of $X_l$. The matrix $M'$ is updated using $M''$, so that
it represents reachabilities between each pair of vertices in $X_l$ when the graph is confined
to $X_l$ and the subtrees rooted at $X_a$ and $X_b$. This process is continued till all the children
of $X_l$ are processed. The matrix $M'$ at this stage reflects the pairwise reachabilities between
vertices of $X_l$, when the graph is confined to the subtree rooted at
$X_l$. Note that the storage required while making a recursive call is
only the current reachability matrix $M'$. Recall that $M'$ contains the pairwise
reachabilitities among the vertices in $X_l$ in the subgraph
corresponding to $X_l$ and the subtrees rooted at those children of
$X_l$ which are processed so far. 
We give the complexity analysis in the
full version.
\begin{lemma}\label{lem:ktrc}
The procedure described above can be implemented in log-space.
\end{lemma}
%\begin{proof}
%It is easy to see that Step $1$ can be implemented in log-space. For Step $2$,
%a log-space transducer computes the reachability matrices $M'$ for each node $X_l$ on $\rho$. 
%Then the log-space algorithm for $k$-paths is invoked. 
%
%%We now show that the reachability matrix described above 
%%for a node $X_l$ 
%%on $\rho$ can be computed in log-space. 
%Now consider the computation of $M'$. While making a recursive call for reachabilities in the child 
%$X_a$ of $X_l$, we do not store
%anything on stack. On the other hand, while making a recursive call for
%reachabilities in subsequent children of $X_l$, we store the matrix $M'$ constructed so far. 
%As $k$ is constant and the size of all the subtrees rooted at the children of $X_l$ excluding $X_a$ 
%is at most $\frac{N}{2}$, the stack depth is always $O(\log{n})$ and thus the algorithm works in log-space.
%\end{proof}
%This matrix is stored on stack
%and reachabilities for another child $X_b$ of $X_l$ within its subtree are computed. 
%and a similar reachability matrix is obtained for 
%\end{enumerate}
%\begin{lemma}\label{lem:ktrc}
%The procedure described above can be implemented in log-space.
%\end{lemma}
\paragraph{Hardness for \Log:}
\Log-hardness of reachability in $k$-trees follows from \Log-hardness of the problem of path ordering
(proved to be \SL-hard in \cite{Ete97}, and is \Log-hard due to \SL=\Log\ result of \cite{Reingold}).
We give the details in the full version.
%\begin{definition} 
%(Path Ordering) {\sc ORD}$(P,u,v)$ Given a directed path $P$, specified by giving for each 
%vertex $w$ (except the last) its successor $s(w)$ along the path, and given vertices $u$, $v$, decide if $u$ precedes $v$
% in $P$.  
%\end{definition}
%As a directed path is a $k$-path for $k=1$, the hardness of $1$-path follows.
%It is also not difficult to see that for any $k\geq 1$ {\sc ORD} is \ACz\ many-one reducible to $k$-path reachability.
%This completes the proof of Theorem \ref{thm:reach}.

%% file: 4-shortlong.tex
\section{Shortest and Longest Paths}\label{sec:path}
We show that the shortest and longest paths in
weighted directed acyclic $k$-trees can be computed in
log-space, when the weights are positive and are given in unary. 
Throughout this section, the terms $k$-path and $k$-tree always refer to directed acyclic
$k$-paths and $k$-trees respectively, with integer weights on edges and 
we here onwards omit the specification \textit{weighted directed acyclic}. 
We use the following (weighted) form of the result from \cite{LMN09}:
The proof is exactly similar to that in \cite{LMN09} and we omit it here.
\begin{theorem}[See\cite{LMN09}, Theorem $9$]
  \label{thm:jt-ext}
  Let $\Class$ be any subclass of weighted directed acyclic graphs closed under
  vertex deletions. 
  There is a function $f$, computable in log-space with oracle access to
  $\Reach(\Class)$, that reduces $\Distance(\Class)$ to
 $\LongPath(\Class)$ and $\LongPath(\Class)$ to $\Distance(\Class)$,
  where ${\Reach(\Class)}$, $\Distance(\Class)$, and $\LongPath(\Class)$
  are the problems of deciding reachability, computing distance and
  longest path respectively for graphs in $\Class$.
 \end{theorem}

We use this theorem to reduce the shortest path problem in $k$-trees to the longest path problem, and then
compute the longest (that is, maximum weight) $s$ to $t$ path. The reduction involves changing
the weights of the edges such that the shortest path becomes the longest path and 
vice versa. This gives a directed acyclic $k$-tree with positive integer weights on edges
given in unary. The class of $k$-trees is not closed under
vertex deletions. However, once a tree decomposition of a $k$-tree is
computed, deleting vertices from the cliques leaves some cliques of size smaller
than $k$, which does not affect the working of the algorithm.

We show that the maximum weight of an $s$ to $t$ path can be computed in
log-space using a technique which uses ideas from \cite{JT07}. The algorithm to compute maximum weight 
$s$ to $t$ path in $k$-trees uses the algorithm for computing maximum weight path in $k$-paths as subroutine.
Therefore we first describe the algorithm for $k$-paths in Section \ref{sec:kpmw}

\subsection{Maximum Weight Path in Directed Acyclic $k$-paths}\label{sec:kpmw}
Let $G$ be a directed acyclic $k$-path and $s$ and $t$ be two designated vertices in $G$.
The computation of maximum weight of an $s$ to $t$ path is done in five stages, described below in detail.
The main idea is to obtain a log-depth circuit by a suitable modification of Algorithm \ref{alg:isreach}, and
to transform this circuit to an arithmetic formula over integers, whose value is used to compute the maximum 
weight of an $s$ to $t$ path in $G$.

Computing the maximum weight $s$ to $t$ path in $G$ involves the
following steps:
\begin{enumerate}
\item {\bf Construct a log-depth formula from Algorithm \ref{alg:isreach}:}
 Modify Algorithm \ref{alg:isreach} so that it outputs a circuit $\mathcal{C}$ that has nodes corresponding to
 the recursive calls made in Line $15$
and the tuples considered in the \textbf{for} loop in Line $14$.
%We use this 
%circuit in the later steps for the computation of maximum weight of an $s$ to $t$ path.
 %we prove that
%this circuit has depth $O(\log{n})$ and hence it can be converted to a log-depth formula $\mathcal{F}$ with only polynomial blow-up in
%size. 
%This circuit is of depth $O(\log{n})$ and 
%hence can be considered as a formula. 
%The nodes of the circuit 
%correspond to the recursive calls made by Algorithm \ref{alg:isreach} and the tuples considered in Line $14$ of
%Algorithm \ref{alg:isreach}. 
%with designated nodes $s$ and $t$ 
%and output a tree which represents the computation structure involving
%the recursive calls and the tuple selection of the algorithm.% consisting of the recursive calls
% . This corresponds to the computation-tree traversed
% by the reachability algorithm. 
A node $q$ in $\mathcal{C}$ that corresponds to a recursive call
 IsReach($u$, $v$, $i$, $j$) has children $q_1,\cdots,q_N$, which correspond to the tuples
considered in that recursive call (\textbf{for}-loop on Line $12$ of Algorithm \ref{alg:isreach}). 
We refer to $q$ as a \textit{call-node} and $q_1,\ldots, q_N$ as \textit{tuple-nodes}. 
%$N$ is the number of tuples considered
%in the recursive call corresponding to $q$.
A tuple-node $q'$ corresponding to a tuple $(v_1,\ldots,v_N)$ 
has call-nodes $q_1',\ldots,q_N'$ as its children,
which correspond to the recursive calls made while considering the tuple
$(v_1,\ldots,v_N)$ (Line $15$ of Algorithm \ref{alg:isreach}). The leaves of $\mathcal{C}$ 
are those recursive calls which satisfy the \textbf{if} condition on Line $3$ of Algorithm \ref{alg:isreach}, 
thus they are always call-nodes. As the depth of the recursion in Algorithm \ref{alg:isreach} is $O(\log{n})$,
the circuit $\mathcal{C}$ also has $O(\log{n})$ depth. Hence it can be converted to a formula $\mathcal{F}$ by only a polynomial
factor blow-up in its size. The maximum number of children of a node is $O(k^k)$ and hence the size of $\mathcal{F}$ is bounded
by $O(k^{k\log{n}})$, which is polynomial in $n$ for constant $k$.
%We give the proof of the following lemma in appendix. 
%\begin{lemma}\label{lem:for}
%The circuit is of depth $O(\log{n})$ and of size $n^{O(k\log{k})}$.
%\end{lemma}

\item {\bf Prune the boolean formula:}
%As the circuit is of depth $O(\log{n})$, it can be converted to a formula with only polynomial increase in its
%size. 
The internal call-nodes of $\mathcal{F}$ are replaced by $\lor$ gates and tuple-nodes are replaced by $\land$ gates.
The leaves of $\mathcal{F}$ are replaced by $0$ or $1$ depending on whether the corresponding recursive call returned $0$
or $1$ in the \textbf{if} block on Line $3$ of Algorithm \ref{alg:isreach}. It can be seen that a sub-formula of $\mathcal{F}$
rooted at a call-node evaluates to $1$ if and only if the corresponding recursive call returns $1$ in Algorithm \ref{alg:isreach}.
Similarly, the sub-formula rooted at a tuple-node evaluates to $1$ if and only if the conjunction corresponding to it (on 
Line $15$ of Algorithm \ref{alg:isreach}) evaluates to $1$. 
 %  and $\lor$ labels to 
% internal nodes, depending on whether an AND or OR operation is
% performed at each node.
%Notice that the leaves always correspond to IsReach calls. Assign
%$0$ or $1$ labels to leaves, depending on whether they return a $0$ or
%a $1$ answer. The tree will be of $O(\log n)$ depth since each time
%the path length in the next recursive IsReach call is halved.
Now, we evaluate the sub-formula rooted at each node of $\mathcal{F}$. 
Note that a node that evaluates to $0$ does not contribute
to any path from $s$ to $t$, and hence its subtree can be safely removed. 

\item  {\bf Transformation into a $\{+,max\}$-tree:} The new, pruned formula obtained in Step $2$ is then relabeled: Each
$\land$ label is replaced with a $+$ label and each $\lor$ label with a 
$max$ label. Each leaf corresponds to calls of the form
$IsReach(u,v,i,i+1)$. It is labeled with the length of the maximum weight
$u$ to $v$ path confined within cliques $i$ and $i+1$, which can be computed in $O(1)$ space. 
This weight is strictly positive, since the $0$-weight leaves are removed in Step $2$. Further, all
the weights are in unary. Thus we now have a $\{+,max\}$-tree $T$ with
positive, unary weights on its leaves. 
%We give the proof of the following simple lemma in the appendix:
%\begin{lemma}\label{lem:+max}  
It is easy to see that the value of the $\{+,max\}$-tree $T$ is the maximum weight of any $s$ to $t$ path in $G$.
%\end{lemma}

\item {\bf Transformation into a $\{+,\times \}$-tree: }
The evaluation problem on the $\{+,max\}$-tree $T$ obtained in Step $3$ is then reduced to the evaluation problem on a $\{+,\times\}$-tree
$T'$ whose leaves are labeled with positive integer weights coded in binary. 
This reduction works in log-space and is similar to that of \cite{JT07}. 
The reduction involves 
replacing a $+$-node of $T$ with a $\times$-node, and a $max$-node with a $+$ node. 
The weight $w$ of a leaf is replaced with $r^{mw}$, where
$r$ is the smallest power of $2$ such that $r\geq n$, and 
$m$ is the sum of the weights of all the leaves of $T$ plus one.
The correctness of the reduction follows from a similar result in
\cite{JT07}, and we omit the proof here.
%in the following lemma. The proof is given in the appendix.
%\begin{lemma}\label{lem:red} 
%The evaluation problem on the $\{+,max\}$-tree with positive integer weights on its leaves, encoded in unary can be reduced in log-space to 
%the evaluation problem on a $\{+,\times\}$-tree with positive integer weights on its leaves, encoded in binary.
%\end{lemma}
%\begin{Proof}[Sketch]
%Let $T$ be a $\{+,max\}$-tree and let $m$ be the sum of weights of all the leaves plus one.
%For each leaf, we replace its weight $w$ by 
\item {\bf Evaluation of the $\{+,\times \}$ tree:} This can be done in
log-space due to \cite{BCGR92, BOC92, CDL01, Hes01}.
The value of $T$ is $v=\lfloor \frac{log_rv'}{m}\rfloor$.
\end{enumerate}

\subsection{Maximum Weight Path in Directed Acyclic $k$-trees}
Given a directed acyclic $k$-tree (in its tree-decomposition) $G$, two vertices $s$ and $t$ in $G$, and weights on the edges of $G$,
encoded in unary, we show how to compute the maximum weight of an $s$ to $t$ path in $G$.
Unlike the case of $k$-paths, the reachability algorithm for $k$-trees given in Section \ref{sec:ktreach} 
can not be used to get a log-depth circuit since the recursion depth of the algorithm
is same as the depth of the $k$-tree. Therefore we need to find another way of recursively dividing
the $k$-tree into smaller and smaller subtrees, as we did for $k$-paths in Sections \ref{sec:kpreach} and \ref{sec:kpmw}.
This is based on the technique used in the following result of \cite{LMR07}:
\begin{lemma}\label{lem:lmr}({\bf Lemma $6$ of \cite{LMR07}, also see \cite{BV83}})
Let $M$ be a visibly pushdown automaton accepting well-matched strings over an alphabet $\Delta$.
Given an input string $x$, checking whether $x\in L(M)$ can be done in log-space.
\end{lemma}
Using Lemma \ref{lem:lmr}, we can compute \textit{a set of recursive
separators for a tree} defined below:
\begin{definition}\label{def:rec}
Given a rooted tree $T$, separators of $T$ are two nodes $a$ and $b$ of $T$ such that
%\begin{enumerate}\label{def:rec}

1. The subtrees rooted at $a$ and $b$ respectively are disjoint,

2. $T$ is split into subtrees $T_1$, $T_2$, $T_3$ where $T_1$ consists of $a$, some (or possibly all) of the children of $a$,
and subtrees rooted at them, $T_2$ is defined similarly for $b$, and $T_3$ consists of the rest of the tree along with a copy of $a$ and $b$ each.

3. Each of $T_1$, $T_2$, $T_3$ consists of at most a $\frac{3}{4}$ fraction of the leaves of $T$.
%\end{enumerate}

This process is done recursively for $T_1$, $T_2$, $T_3$, until the number of leaves
in the subtrees is two. Such a subtree is in fact a path.
A \textit{set of recursive separators of $T$} consists of the separators of $T$ and of all the subtrees obtained in the
recursive process.
\end{definition}
The following lemma gives the procedure to compute a set of recursive separators of a tree $T$:
\begin{lemma}\label{lem:ktd}
Given a tree $T$, the set of recursive separators of $T$ can be computed in log-space. 
\end{lemma}
\begin{proof}
%We do not go into the details of visibly pushdown automaton here, but instead, we describe how 
%this algorithm can be used to get the desired splitting procedure for $k$-trees. 
The algorithm of \cite{LMR07} deals with well-matched strings. An example of a well-matched string is a balanced
parentheses expression, which is a string over $\{(,)\}$. 
In \cite{LMR07}, a log-space algorithm is given for membership testing
in those languages which are subsets of 
well-matched strings and are accepted by visibly pushdown automata. 
%Subsets of balanced parentheses expressions 
We restrict ourselves to balanced parentheses expressions.
 To check whether a string on parentheses is in the language, the algorithm
of \cite{LMR07} recursively partitions the string into three disjoint substrings, such that each of the parts 
forms a balanced parentheses expression, and length of each part is at most $\frac{3}{4}$th of the length of the original string.
To use this algorithm, we order the children of each node of $T$ in a specific way,
label the leaves with parentheses $`('$ and $`)'$ such that the leaves of the subtree
rooted at any internal node form a string on balanced parentheses. We add dummy leaves if needed.
The steps are as follows:
%As in the case of $k$-paths, the main idea here is to divide the tree 
%recursively into three subtrees. Each of these subtrees formed at step $i$ contain a 
%constant fraction of the number of leaves of 
%the subtree in step $i-1$, from which they are formed. 
%The subtrees share only one 
%clique. Therefore an $s$ to $t$ path can pass through this clique at most $k$ 
%times. Thus, 
% This circuit is an arithmetic circuit of log-depth, and hence can 
%be converted to a tree. The operations in this tree are $\{+,max\}$. 
%Exactly as in $k$-paths case, we assign appropriate values to the leaves of this
%tree, and convert it to a $\{+,\times\}$ tree. All these operations can be done in
%log-space, and the resulting arithmetic formula can also be evaluated in log-space.
%For the recursive division into subtrees, we use the log-space algorithm of 
%\cite{LMR07}. This algorithm takes as input an expression on $`('$ and $`)'$ and 
%checks in log-space whether the expression has balanced parentheses. Hence we 
%assign $`('$ and $`)'$ labels to the leaves of the $k$-tree and use this algorithm.
%The main steps are as follows:
%\begin{enumerate}

1. By adding dummy leaves, ensure that each internal node has an even number of children
which are leaves, and there are at least two 
such children. 

2. Arrange the children of each node from left to right such that the non-leaves 
are consecutive, and they have an equal number of leaves 
to the left and to the right.

3. For each internal node, label the left half of its leaf-children with `(' and the right ones
by `)'. This ensures that the leaves of the subtree rooted at each internal node form a 
balanced parentheses expression. Conversely, leaves which
form a balanced parentheses expression are consecutive leaves in the subtree rooted at an
internal node.
%Assign `(' or `)' to each of the leaves such that the labels of the leaves of the 
%subtree rooted at each internal node form a string of balanced parentheses. 
%As each internal node
%This is done by assigning the label `(' (respectively `)') to the left most (right most) leaf in a subtree. 
%Thus, for an internal node, starting from the leftmost child which is a leaf, we 
%assign `(' and `)' alternately. If there are one or more non-leaf children, there 
%is a leaf-child preceding them and following them. These leaves get `(' and `)' 
%respectively. Thus the group of non-leaf children of a node is enclosed in a pair of
%`(' and `)'. 
%\end{enumerate}

The leaves of $T$ now form a balanced parentheses expression, and we run the algorithm of \cite{LMR07} on this
string. The recursive splitting of the string into smaller substrings corresponds to the
recursive splitting of $T$ at some internal nodes, which satisfies Definition \ref{def:rec}. 
This is ensured by the way the leaves are labeled. Each balanced parentheses expression corresponds to
either a subtree rooted at an internal node or the subtrees rooted at some of the children of an internal node.

The subtrees obtained by splitting a tree have at most $\frac{3}{4}$th of the number of leaves in the tree.
Thus at each stage of recursion, the number of leaves in the subtrees is reduced by a constant fraction. 
%When
%there are only two leaves in a subtree, the recursion stops. 
Moreover, the algorithm of \cite{LMR07} can output all the substrings formed at each stage of recursion in log-space. 
As a substring completely specifies a subtree of $T$, our procedure outputs the set of recursive separators for $T$ 
in log-space.
\end{proof}

Once an algorithm to compute the set of recursive separators for $k$-trees is known,
a reachability routine similar to Algorithm \ref{alg:isreach} can be designed in a straight forward way. We give the details in the full version.
From the reachability routine, the computation of maximum weight
path follows from the steps $1$ to $5$ described in Section \ref{sec:kpmw}.
%Therefore we get the following theorem:
%\begin{theorem}
%Given a (weighted) directed acyclic $k$-tree $G$ with unary weights on edges, 
%and two designated vertices $s$ and $t$ in $G$, the maximum weight of
%a simple path from $s$ to $t$ can be computed in log-space.
%\end{theorem}
%Using this, we recursively find a \emph{valid} substring
%of this string represented by all the leaves of the tree. The corresponding leaves 
%have a common ancestor, say $C$, where the original tree is split into three 
%subtrees. We refer to \cite{LMR07} for details. Each subtree has a 
%constant fraction of the number of leaves of the parent subtree. This can be done 
%in log-space. The recursion depth is $O(\log{n})$, as the number of leaves in a 
%subtree is reduced by a constant fraction in each step.
%Suppose the tree is split at a clique $C$. Then $C$ is the only node which 
%appears in all the three subtrees. Thus a path from $s$ to $t$ can pass through $C$
%at most $k$ times. Similar to $k$-paths case, we cycle through all the $k$ tuples 
%of the vertices in $C$, to check whether $t$ is reachable from $s$. The separated 
%node $C$ can be recomputed each time we return from the recursive call, by 
%the log-space algorithm of \cite{LMR07}.
%The base case is when we have two leaves. In this case, we need to consider 
%only the path between the two leaves and thus $k$-path reachability algorithm can
%be used.
\subsection{Distance Computation in Undirected $k$-trees}
We give a simple log-space algorithm for computing the shortest 
path between two given vertices in an undirected $k$-tree. 
We use the decomposition of 
\cite{KCP82}, where a $k$-tree is decomposed into layers. We use the following 
properties of the decomposition:
%\begin{itemize}

1. Layer $0$ is a 
$k$-clique. Each vertex in layer $i>0$ has exactly $k$ neighbors in layers 
$j<i$. Further, these neighbors of $i$ which are in layers lower than that of
$i$ form a $k$-clique.

2. No two vertices in the same layer share an edge.
%\end{itemize}

This decomposition is log-space computable \cite{KK09}. Moreover, given two
vertices $s$ and $t$, it is always possible to find a decomposition in which
$t$ lies in layer $0$. This can also be done in log-space. 
If both $s$ and $t$ are in layer $0$, then there is an edge between $s$ and $t$,
which is the shortest path from $s$ to $t$. Therefore assume that $s$ lies in
a layer $r>0$. 
The following claim leads to a simple algorithm. The proof appears
in the full version.
\begin{claim}\label{clm:1}
1. The shortest $s$ to $t$ path never passes through two vertices $u$ and $v$ such
that $layer(u)<layer(v)$. 
%\end{claim}
%\begin{claim}\label{clm:2}
2. There is a shortest path from $s$ to $t$ passing through the 
neighbor of $s$ in the lowest layer.
\end{claim}
This claim suggests a simple algorithm which can be implemented 
in log-space: Start from $s$ and choose the next vertex from the lowest possible
layer, at each step till we reach layer $0$.

%% file: 5-matchings.tex
\section{Perfect Matching in $k$-trees}\label{sec:mat}
\paragraph{\bf Hardness for \Log:}
To show that the decision version of perfect matching is
hard for \Log, we show that the problem of path ordering,
%defined in Section \ref{sec:reach} 
can be reduced to the perfect matching 
problem for $k$-trees. We give the proof in the full version:
%\begin{lemma}\label{lem:mhard} (see also \cite{DR05})
%Determining whether a tree has a perfect matching is hard for \Log
%under projections. 
%\end{lemma}
%Since a tree is a $k$-tree with $k = 1$ we get:
\begin{lemma}\label{cor:hard}
 Determining whether a $k$-tree has a perfect matching is $\Log$-hard.
\end{lemma}
\paragraph{\bf \Log\ upper bounds:}
We describe a log-space algorithm to decide whether a $k$-tree 
has a perfect matching and, if so, output a perfect matching. 
The algorithm is inspired by an $O(n^3)$ algorithm \cite{CH92}
for computing the matching polynomial in series-parallel graphs.
The idea is to exploit the fact that 
$k$-trees have a tree decomposition of bounded width, so that
any perfect matching of the entire $k$-tree induces a partial matching
on any subtree which leaves at most constantly many vertices unmatched.
Thus we generalize the problem to that of determining,
for each set, $S$, of constantly many vertices in the root of the subtree, 
whether there is a matching of the subtree that leaves exactly the vertices
in $S$ unmatched. Now we ``recursively'' solve the generalized problem
and for this purpose we need to maintain a bit-vector indexed by 
the sets $S$ which is still of bounded length. The algorithm composes
the bit-vectors of the children of a node to yield the bit-vector for the node.
The bit-vector, which we refer to as \textit{matching vector}, is defined as follows:
\begin{definition}
Let $G$ be a $k$-tree with tree-decomposition $T$. $T$ has alternate levels of $k$-cliques and
$k+1$-cliques. Root $T$ arbitrarily at a $k$-clique. Let $s$ be a node in $T$ that shares
vertices $\{u_1,\ldots,u_k\}$ with its parent. Further, let $H$ be the subgraph of $G$ corresponding
to the subtree of $T$ rooted at $s$. The \textit{matching
vector} for $s$ is a vector $\vec{v}_H=(v_H^{(S_1)},\ldots ,v_H^{(S_{2^k})})$ of dimension $2^k$, where
%\begin{itemize}
$S_1,\ldots ,S_{2^k}$ are all the distinct subsets of $\{u_1,\ldots ,u_k\}$,
and $v_H^{(S_i)}=1$ if $H$ has a matching in which all the vertices of $H$ matched, except those in $S_i$,
$v_H^{(S_i)}=0$ if there is no such matching.
%\end{itemize}
\end{definition}
It can be seen that $G$ has a perfect matching if and only if
$v_G^{(\emptyset)}=1$.%, where $\emptyset$ is the empty set.
We show how to compute $\vec{v}_G$ in \Log, and also show how to construct a perfect matching in $G$, if one exists.
We prove Part 1 of the following theorem. For a proof of part 2, we
refer to the full version.
\begin{theorem}\label{thm:ub}
%\begin{enumerate}
1. The problem of deciding whether a $k$-tree has a perfect matching is in \Log.

2. Finding a perfect matchings in a $k$-tree is in \FL.
%\end{enumerate}
\end{theorem}
\begin{proof}({\bf of $1$})
We compute the matching vector for the root by recursively
computing the matching vectors of each of its children.
For a leaf node in the tree-decomposition, the matching vector can be computed in a brute-force way.
At an internal node $s$, the matching vector is computed from the matching vectors of its children, which
we describe here:

{\bf Case $1$: $s$ is a $k$-node} Let $s$ has vertices
$V_s = \{u_1,\ldots,u_k\}$. Recall that a $k$-node shares all its vertices with all its neighbors.
Let the children of $s$ in $T$ 
be $s_1,\ldots,s_r$. Let the subgraph corresponding to the subtree rooted at $s$ be $H$ and those at
its children be $H_1,\ldots,H_r$. 
%% The intuitive idea behind this composition of matching vectors of the children
%% is as follows: $H$ has a matching with a subset of vertices $S_i\subseteq \{v_1,\ldots,v_k\}$ unmatched
%% and all other vertices matched if and only if $H_1,\ldots,H_r$ have a matching with $S_i$ unmatched
%% and each of the vertices in $\{v_1,\ldots,v_k\}\setminus S_i$ are matched in exactly one of the subgraphs $H_1,\ldots,H_r$.
In order to determine $v_H^{(S)}$, we need to know if there is a matching in $H$ that leaves
 exactly the vertices in $S$ unmatched. This holds if and only if the vertices in $S$ are not matched in any of the $H_j$'s,
and each vertex in $V_s\setminus S$ is matched in exactly one of the $H_j$'s. In other words,
we need to determine if there is a partition $T_1,T_2,...,T_r$ 
%(which are pairwise disjoint, and together, include exactly the vertices in $V_s \setminus S$) there
of $V_s\setminus S$, such that $H_j$ has a matching in which precisely $V_s\setminus T_j$ is unmatched. That is,
 $v_{H_j}^{(V_s\setminus T_j)}=1$ for all $1\leq j\leq r$.
%each $H_j$, $1\leq j\leq r$ has a matching, that matches all the vertices in 
%$H_j$ except those in $V_s \setminus T_j$
More formally,
\begin{eqnarray}
v_H^{(S)}  =  \bigvee_{\substack{T_1,\ldots,T_r\subseteq {V_s \setminus S}:\\
                  \forall j\ne j' \in [r] T_j\cap T_{j'}=\emptyset: \\
                 \cup_{j \in [r]}{T_j}=V_s \setminus S }} \bigwedge_{j \in [r]}{v_{H_j}^{(V_s \setminus T_j)}} 
	   & = & \bigvee_{\emptyset  = U_0 \subseteq \ldots \subseteq U_r = {V_s \setminus S}}
                 \bigwedge_{j \in [r]}{v_{H_j}^{(V_s \setminus (U_{j} \setminus U_{j-1}))}}
\end{eqnarray}
where, the second equality follows by defining $U_0 = \emptyset$ and $U_i = \cup_{j \in [i]}{T_j}$ for $i \in [r]$.
The size of the above DNF formula depends on $r$ which is not a constant hence the straightforward
implementation of the above computation would not be in \Log.  However, consider a conjunct
in the big disjunction in the second line above. The $j^{\mbox{th}}$ factor of this conjunct
depends only on $U_j$ and $U_{j-1}$, each of which can be represented by a constant number
($= 2^k$) of bits. Thus, we can iteratively extend $U_{j-1}$ in all possible ways to $U_j$
and use the bit indexed by $V_s \setminus (U_j \setminus U_{j-1})$ in the vector for the child.
How to obtain the vector of the child within a log-space
bound is detailed in the full version.
%proof of Lemma \ref{lem:matlog}.
%% Since $r$ is not a constant, the above composition can not be done in a straight forward way.

{\bf Case $2$: $s$ is a $k+1$ node} The procedure is slightly more complex in this case.
Let $s$ have vertices $\{u_1,\ldots,u_{k+1}\}$. Let the subgraph corresponding to the subtree rooted at $s$ be $H$.
Let $s_1,\ldots ,s_r$ be the children of $s$, with corresponding subgraphs $H_1,\ldots ,H_r$.
Note that $s$ may share a different subset of $k$ vertices with each of its children and with its parent.
Let the vertices $s$ shares with its parent be $\{u_1,\ldots,u_k\}$. Then its matching vector is indexed by
the subsets of $\{u_1,\ldots,u_k\}$, and moreover, $u_{k+1}$ should always be matched in $H$. To compute $\vec{v}_H$, we first extend
the matching vectors of each of its children and make a $2^{k+1}$ dimensional vector $\vec{w}_H$.
% where $H$
%is the subgraph corresponding to the subtree rooted at $s$.
The matching vector $\vec{v}_{H_j}$ of a child $s_j$
of $s$ is extended to the new vector $\vec{w}_{H_j}$ as follows: Let $s_j$ contain $\{u_1,\ldots,u_k\}$. We consider
an entry $v_{H_j}^{(S)}$ of $\vec{v}_{H_j}$. The vector $\vec{w}_{H_j}$ has two entries corresponding to it.
$$w_{H_j}^{(S\cup \{u_{k+1}\})}  =  v_{H_j}^{(S)},\qquad
w_{H_j}^{(S)}  =  \bigvee_{\substack{p\in [k],u_p\notin S,\\ (u_{k+1},v_p)\in E}} u_{H_j}^
{(S\cup \{u_p\})}$$
These new vectors of each of the children can be composed similar to that in the previous case to get
$\vec{w}_H$. To get $\vec{v}_H$, we remove the $2^k$ entries from $\vec{w}_H$ which are indexed on subsets containing
$u_{k+1}$. This vector is passed on to the parent of $s$. The complexity
analysis, and a proof of ($2$) appears in the full version.
\end{proof}